\pdfoutput=1

\documentclass{PoS}

\title{Application of Maximum Entropy Deconvolution to ${\gamma}$-ray Skymaps}

\ShortTitle{Application of Maximum Entropy Deconvolution to ${\gamma}$-ray Skymaps}

\author{\speaker{Susanne Raab} and Ira Jung-Richardt\\
	Erlangen Centre for Astroparticle Physics (ECAP), Universit\"at Erlangen-N\"urnberg, Germany\\
  E-mail: \email{Susanne.Raab@fau.de}}
      
\abstract{%
	Skymaps measured with imaging atmospheric Cherenkov telescopes (IACTs) represent the real source distribution convolved with the point spread function of the observing instrument. Current IACTs have an angular resolution in the order of 0.1$^\circ$ which is rather large for the study of morphological structures and for comparing the morphology in $\gamma$-rays to measurements in other wavelengths where the instruments have better angular resolutions. 
	
	Serendipitously it is possible to approximate the underlying true source distribution by applying a deconvolution algorithm to the observed skymap, thus effectively improving the instruments angular resolution. From the multitude of existing deconvolution algorithms several are already used in astronomy, but in the special case of $\gamma$-ray astronomy most of these algorithms are challenged due to the high noise level within the measured data. 
	
	One promising algorithm for the application to $\gamma$-ray data is the Maximum Entropy Algorithm. The advantages of this algorithm are the possibility to take a priori knowledge into account and that it is an independent approach to previous work, e.g., \cite{Heinz2012} who applied the Richardson Lucy Algorithm to $\gamma$-ray skymaps.
	
	An implementation of the Maximum Entropy Algorithm is provided in the \texttt{MemSys5} software package \cite{memsys}. As this algorithm is very sensitive to various input parameters it is essential to understand their influences. We present a study of the influences of these parameters in order to investigate the applicability of the Maximum Entropy Algorithm for the deconvolution of skymaps in $\gamma$-ray astronomy.}

\FullConference{The 34th International Cosmic Ray Conference,\\
		30 July - 6 August, 2015\\
		The Hague, The Netherlands}
	
\begin{document}

\section{Introduction}

Nature does not reveal its mysteries easily. This principle is demonstrated by e.g.\@ studying morphological structures. Due to the limited angular resolution of detectors, images represent always the source distribution convolved with the point spread function (PSF). In the case of very-high-energy (VHE; >100 GeV) $\gamma$-ray images (skymaps) of imaging atmospheric Cherenkov telescopes (IACTs) the size of the PSF is in the order of 0.1$^\circ$, which is rather large, especially compared to other wavelengths, and thus causes a few challenges.

Hence even a perfect point source exhibits an extent -- however exactly the question whether or not a source is a point source is often of crucial interest, cf.\@ for instance \cite{halo}.

A large PSF also limits the possibilities to study structures within the source, especially in the order of or smaller than the PSF. A paramount example for this is the differentiation whether a supernova remnant (SNR) exhibits a shell type morphology or not.

In addition multiwavelength studies are essential in $\gamma$-ray astronomy to obtain further inside into the processes taking place in astrophysical sources. This study is biased by the fact that the PSF of $\gamma$-ray experiments is often orders of magnitude larger compared to other wavebands. For instance the PSF of the X-ray observatory XMM-\textit{Newton} is two orders of magnitude better than the one of H.E.S.S. To overcome this problem the traditional ansatz is to smooth data from the superior instrument with the PSF from the other one, obviously loosing information by doing this.

One way to diminish these problems is the application of adequate deconvolution algorithms to VHE $\gamma$-ray skymaps to approximately derive the underlying true source distribution, thus effectively improving the instruments angular resolution. From the multitude of existing deconvolution algorithms several are already used in astronomy, but in the special case of $\gamma$-ray astronomy most of these algorithms are challenged due to the high noise level within the measured data. In previous works algorithms like the Richardson-Lucy Algorithm \cite{Heinz2012} have successfully been applied to VHE $\gamma$-ray data. In this work we will report a study about the applicability of the Maximum Entropy Algorithm to VHE $\gamma$-ray data.

The main object of the study, presented in this work, is the investigation of parameter influences on the deconvolution and finding an optimal parameter set. This study is based on simulations. In Section~\ref{sec:sim} the simulation of $\gamma$-ray excess maps is outlined. In the subsequent Section~\ref{sec:maxent} the Maximum Entropy Deconvolution, as used in this work, will be described, followed by an overview over our main results in Section~\ref{sec:results}. To conclude we will give a brief summery and outlook in Section~\ref{sec:sum}.

\section{Simulation of $\gamma$-ray skymaps}
\label{sec:sim}

To simulate a $\gamma$-ray excess map realistic input parameters 
e.g.\@ for the camera acceptance, the background rate and the PSF have to be chosen. In our case templates were obtained from the H.E.S.S. \texttt{Model++} analyses \cite{model} of the supernovae remnant (SNR) RX~J1713.7$-$3946 and the Crab nebula for extended and for point sources, respectively. The main background source in $\gamma$-ray astronomy with Cherenkov telescopes are events from Cosmic rays which are evenly distributed over the field of view. Therefore a homogeneous background distribution is assumed. The number of background events is taken from the analysis mentioned above as well as the number of excess events. The latter one is scalable according to the simulated source significance. As source morphology different source type can be chosen, e.g.\@ point like. 

\begin{figure}[h]
	\centering
	\includegraphics[width=.32\textwidth]{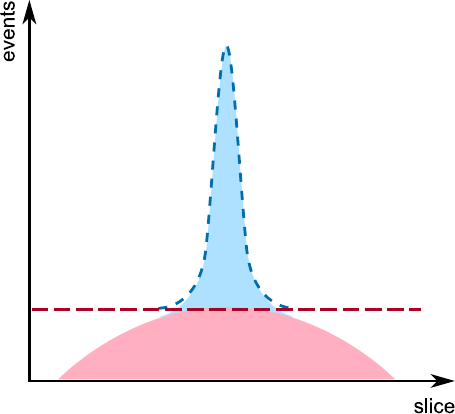}
	\includegraphics[width=.32\textwidth]{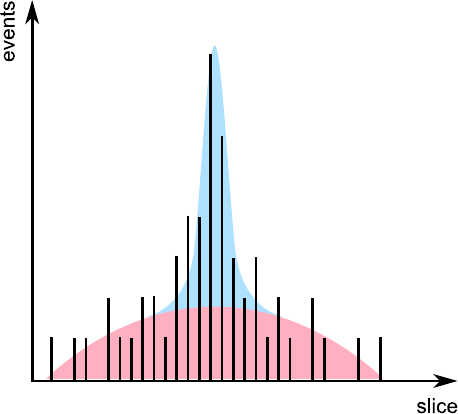}
	\includegraphics[width=.32\textwidth]{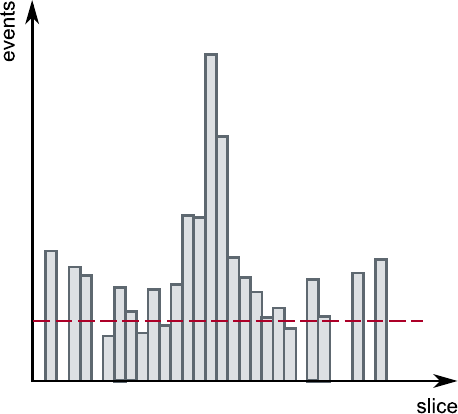}
	\caption{Schematic view of the simulation process (not to scale). \emph{Left panel:} Flat background (dashed red) with a source (dashed blue) on top of it. In solid colours the background and source modulated with the $\gamma$-ray acceptance. \emph{Central panel:} The black bars represent the simulated $\gamma$-ray and background events. \emph{Right panel:} Simulated events corrected for the $\gamma$-ray acceptance.}
	\label{fig:sim}
\end{figure}

The main steps of the simulation process are illustrated in Figure~\ref{fig:sim} and are given below.

\begin{enumerate}
	\item[1)] The background and the source distribution is convolved with the $\gamma$-ray acceptance. The dashed lines in Figure~\ref{fig:sim} (\emph{left panel}) show the unconvolved background (\emph{red}) and source (\emph{blue}) distributions. The solid areas give the convolved ones.
	\item[2)] Random events are generated according to the distribution obtained in the first step. This is illustrated in Figure~\ref{fig:sim} (\emph{central panel}), the black lines illustrate the simulated events. 
	\item[3)] The map obtained in the second step is corrected for the acceptance. As illustrated in Figure~\ref{fig:sim} (\emph{right panel}) the source in the centre is only slightly affected by the correction, but the background distribution resembles a flat distribution with larger fluctuations at the edges of the skymap. 
	\item[4)] In the final step the above assumed background level is subtracted.
\end{enumerate}

\begin{figure}[ht]
	\centering
	\includegraphics[width=.24\textwidth]{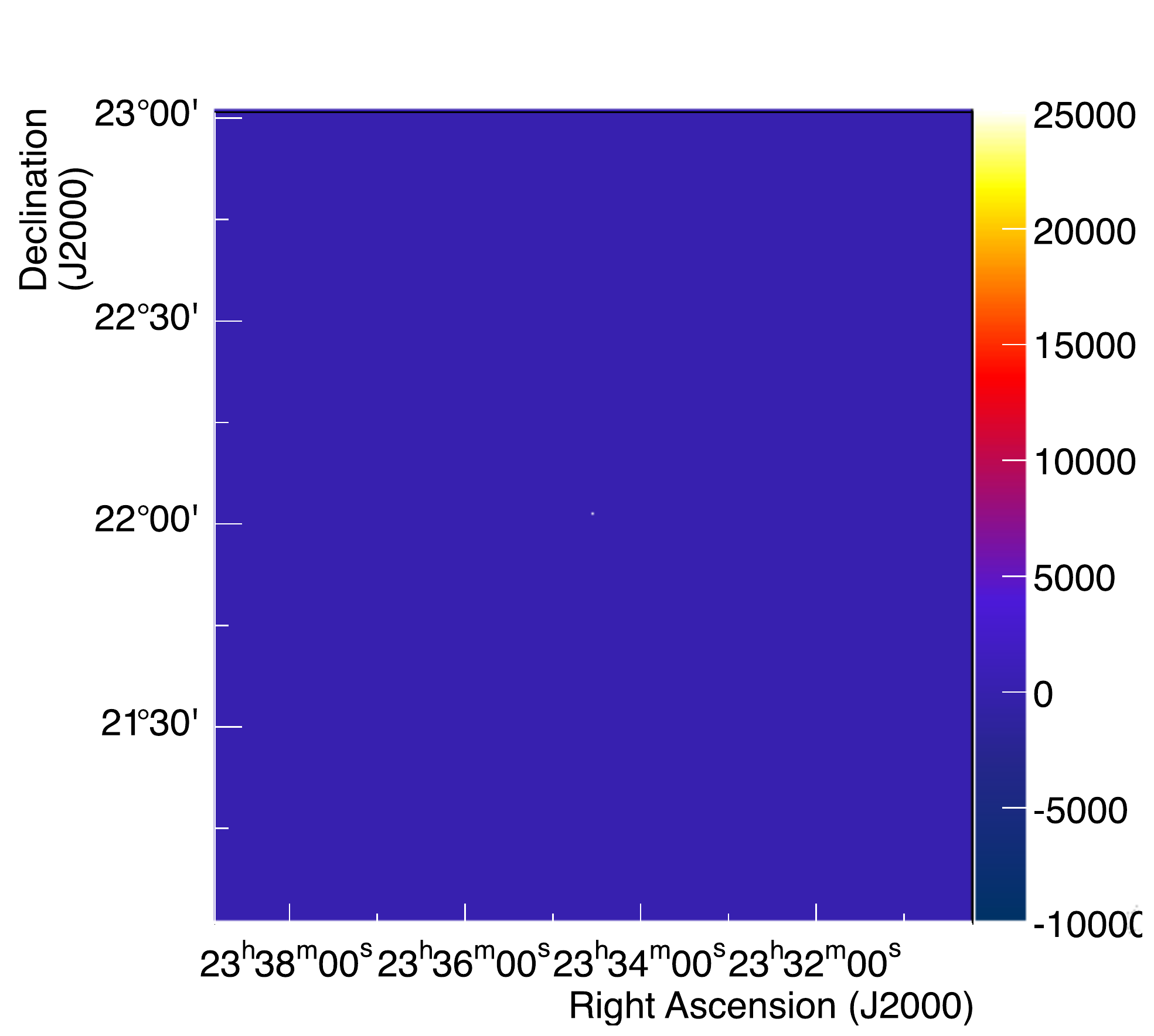}
	\includegraphics[width=.24\textwidth]{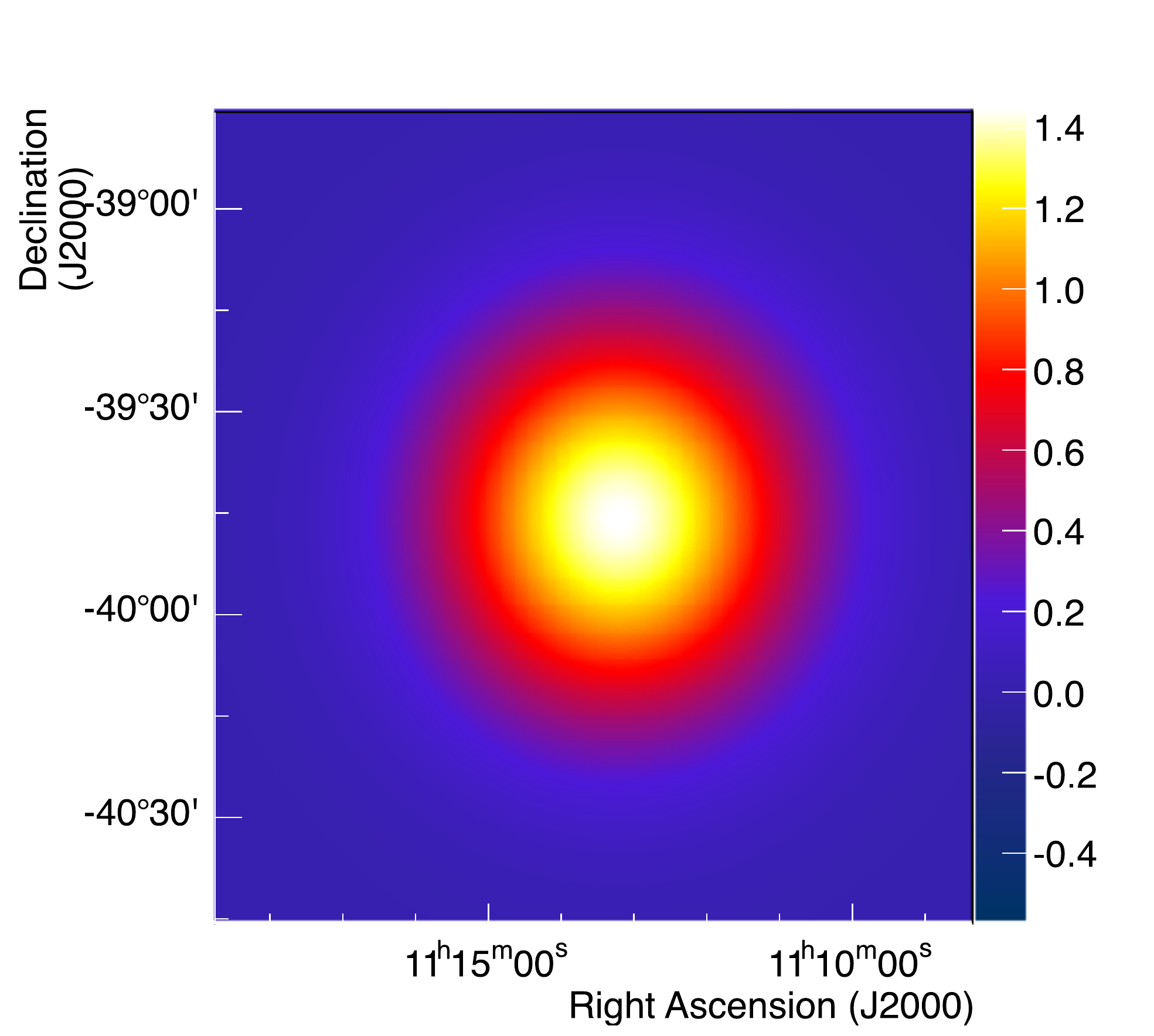}
	\includegraphics[width=.24\textwidth]{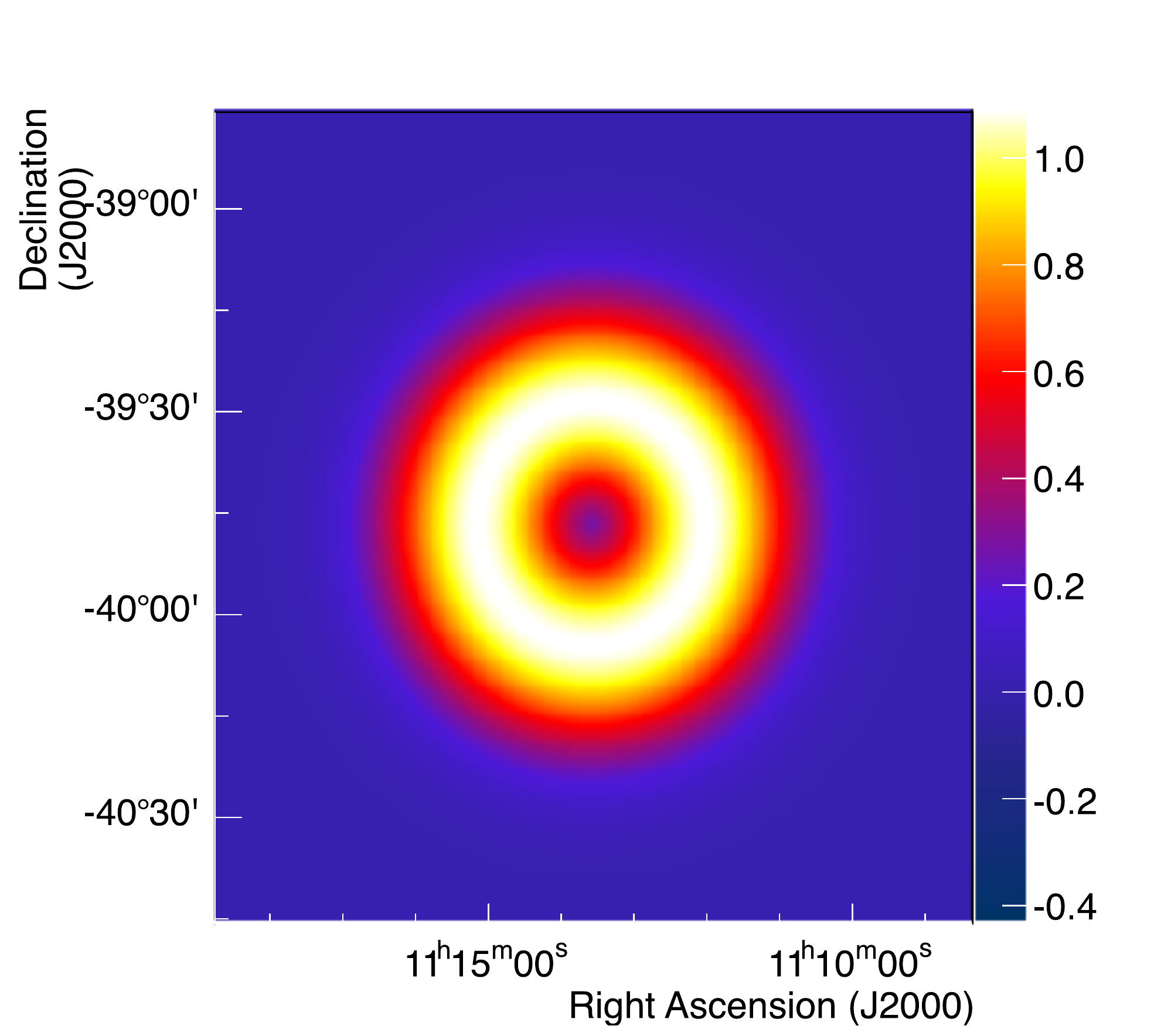}
	\includegraphics[width=.24\textwidth]{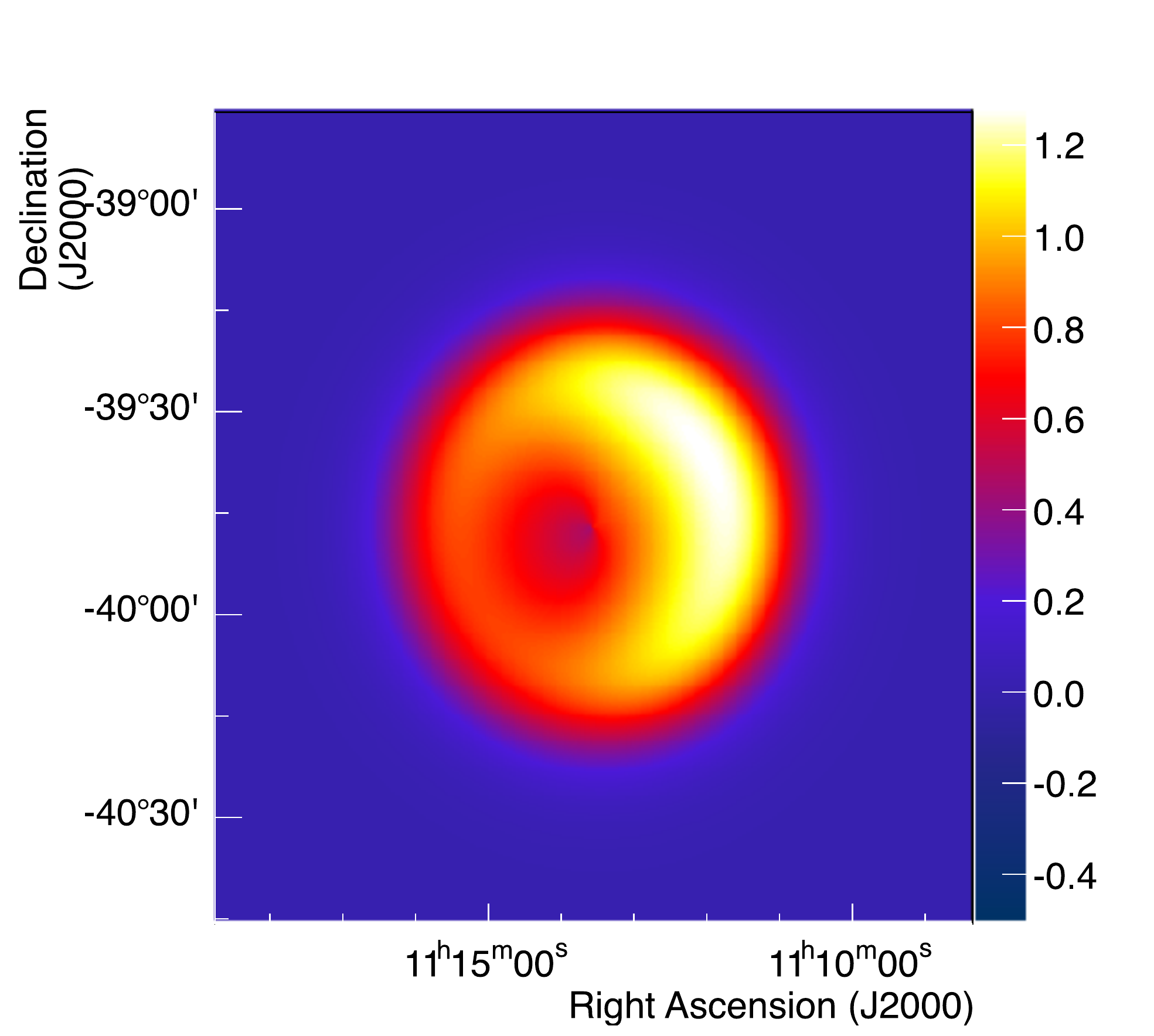}\\
	\includegraphics[width=.24\textwidth]{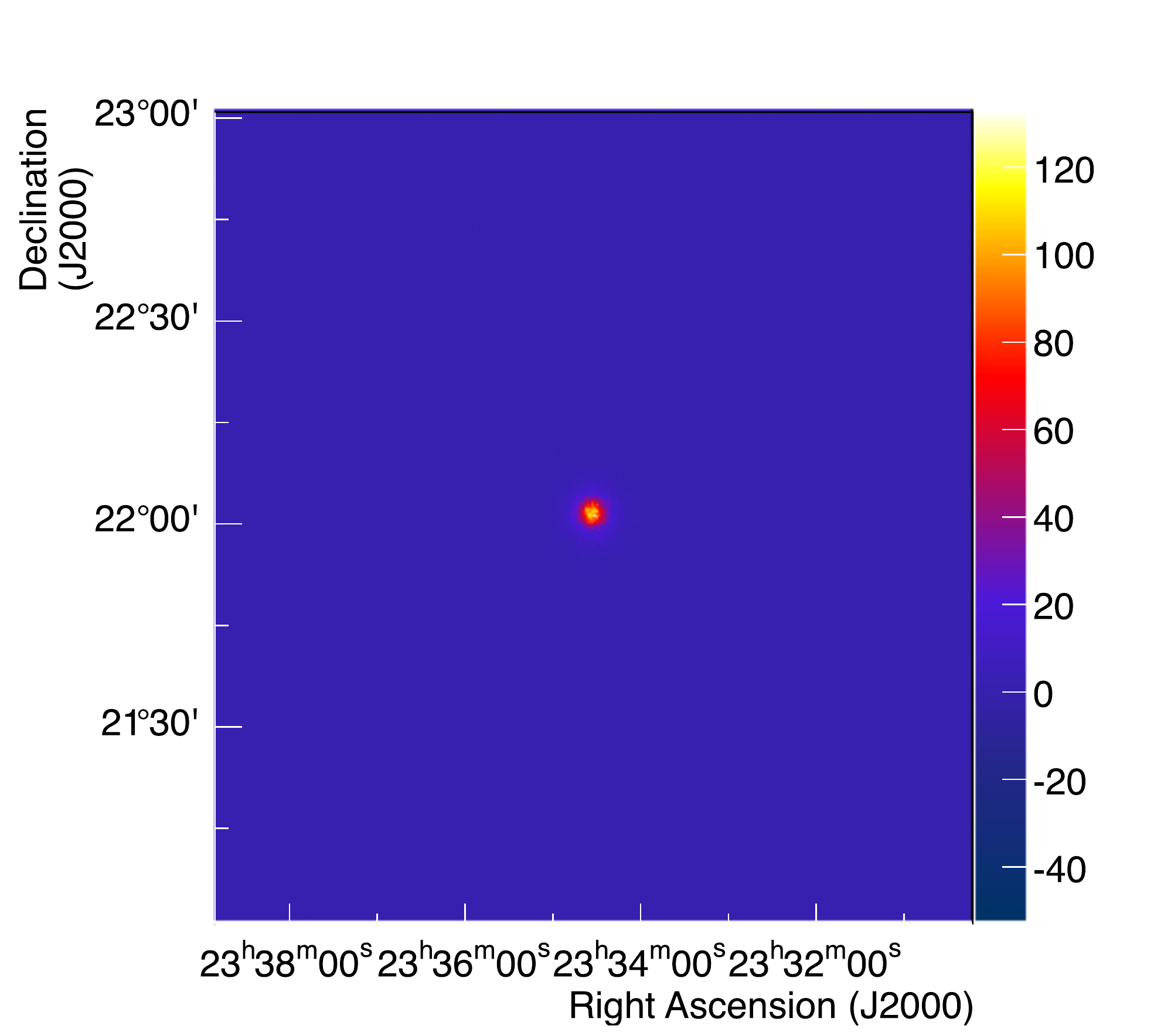}
	\includegraphics[width=.24\textwidth]{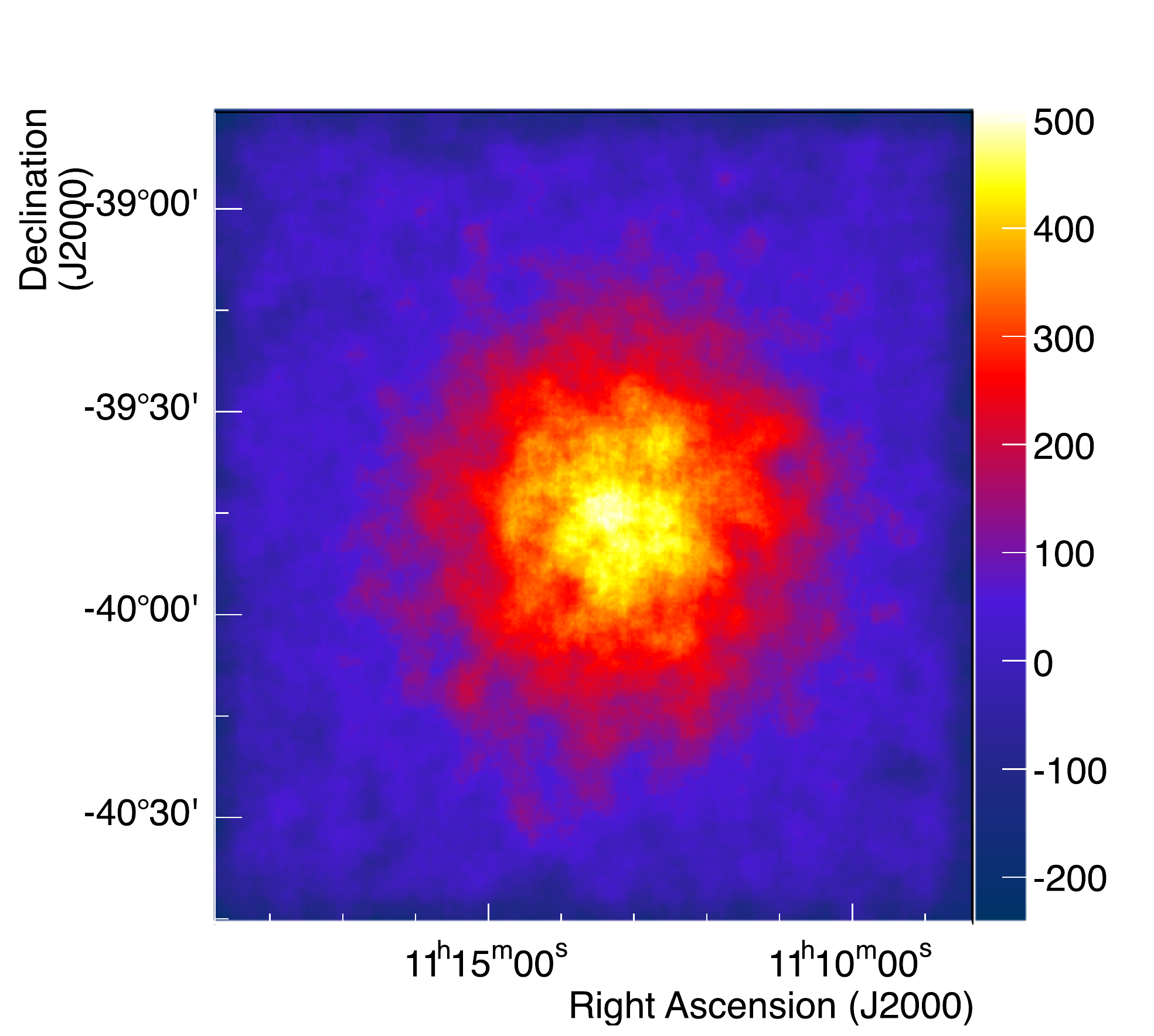}
	\includegraphics[width=.24\textwidth]{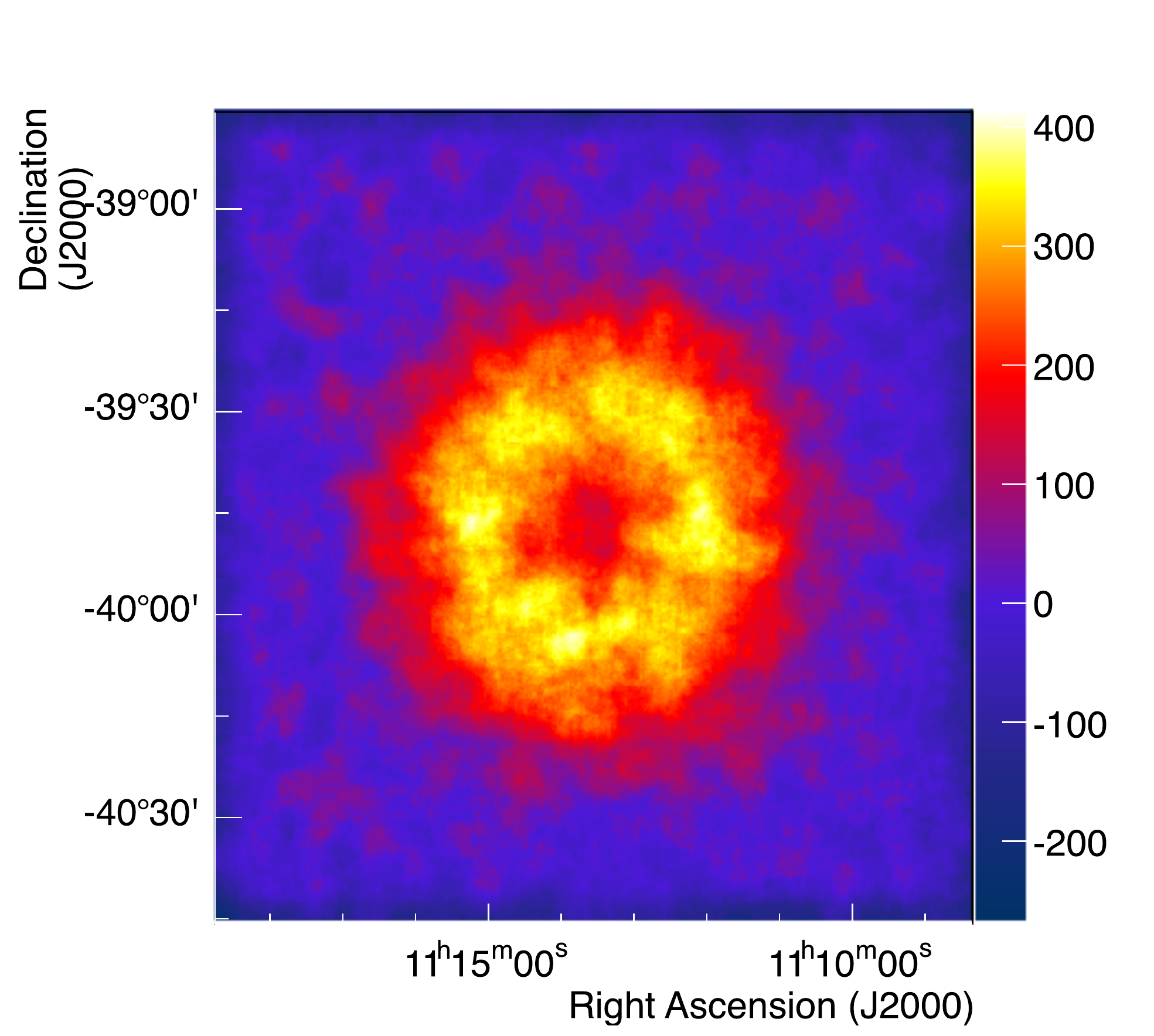}	
	\includegraphics[width=.24\textwidth]{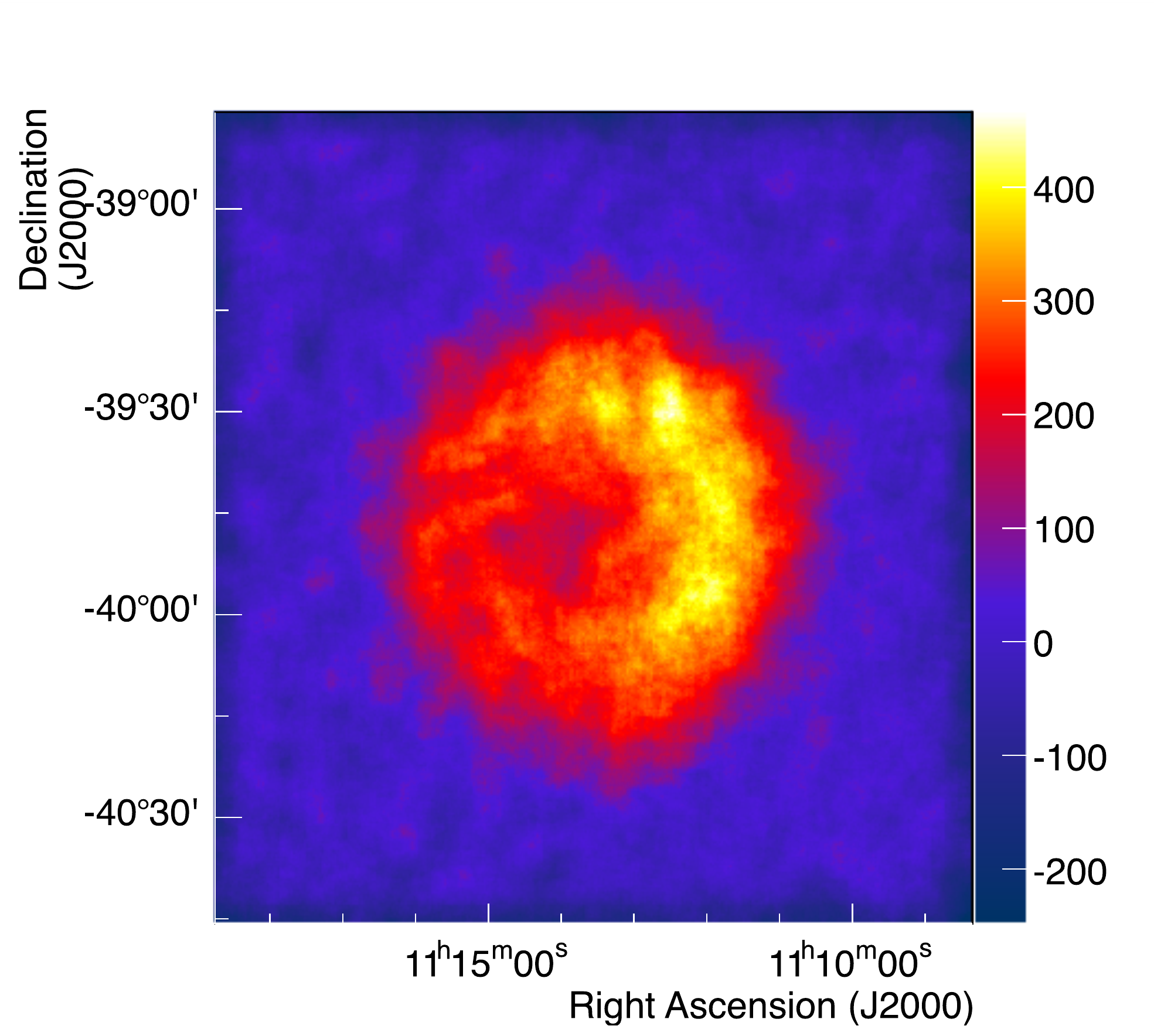}	
	\caption{\emph{Top row:} Examples of source models, which can be used for the simulation. \emph{From left to right:} A point source, a Gaussian, a ring and a modified ring shape, all fitted to resemble the $\gamma$-ray emission of RX~J1713.7$-$3946. \emph{Bottom row:} Simulations according to source shapes displayed in the top row. Besides the point source on the left hand side, the simulations are smoothed with a Gaussian ($\sigma = 0.05^\circ$) for better visual perception.}
	\label{fig:simulations}
\end{figure}

In Figure~\ref{fig:simulations} examples of different source morphologies and simulated skymaps are given. The models of extended sources are adjusted to resemble the $\gamma$-ray emission of RX~J1713.7$-$3946. 

\section{Maximum Entropy Deconvolution}
\label{sec:maxent}

The Maximum Entropy Method \cite{PhysRev.106.620} is widespread in a surprisingly large variety of fields. Its applications range from studying the impact of the climate change on coffee supply chains \cite{coffee} to calculations of drug absorption rates \cite{drugs}. In astronomy it is used in nearly all wavelengths regimes: from radio astronomy (cf., e.g.\@ \cite{1985A&A...143...77C}) to optical images from the Hubble Space Observatory (especially in the period before the refurbishment of the detector system, see, e.g., \cite{1991rhis.conf...31W,1991rhis.conf..132H}) up to the $\gamma$-ray observatory COMPTEL \cite{strong}.

Its basic principle is that the most realistic solution is the one which minimises the amount of information while remaining compatible with the data. This minimisation of information is equivalent to maximising the entropy. From this theory a method was invented to reconstruct images from incomplete and noisy data, cf.\@ \cite{themaxen}. In this work  Maximum Entropy Deconvolution is used in the implementation provided by the \texttt{MemSys5} software package \cite{memsys}.

The deconvolution procedure has several input parameters, which are listed in Table~\ref{tab:parameter}.

\begin{table}[ht]
	\centering
	\begin{tabular}{lp{9.2cm}l}
		\hline
		Parameter & Description & Configurations\\
		\hline
		\texttt{maxiter} & Number of iterations & 1, 2, 4, \dots, 1024\\
		\texttt{nscales} & Number of hidden images, each progressively blurred, to be combined to form the visual image & 7, 8\\
		\texttt{icfwidth} & Width of the intrinsic correlation function & 0, 1, 2, \dots, 8\\
		\texttt{default} & Value of default distribution & 0.001, 0.01, 0.1, 1\\
		\texttt{aim} & Value of the stopping criterion & 0.001, 0.01, 0.1, 1\\
		\texttt{noise} & Standard deviation of the Gaussian noise& 0.0001\\
		\texttt{nsample} & Number of samples taken to calculate standard deviation etc. & 3\\
		\texttt{bayes} & Stopping criterion; ``1'' refers to classic maximum entropy stopping criterion & 1\\
		\texttt{entropy} & Type of entropy; ``1'' refers to standard entropy  & 1\\
		\hline
	\end{tabular}
  \caption{List of parameters used for studying parameter influences on the deconvolution. For an in-depth description of their definition see \cite{memsys}. The configurations denote the tested parameter combinations.}
  \label{tab:parameter}
\end{table}

\begin{figure}[ht]
	\centering
	\includegraphics[width=.45\textwidth]{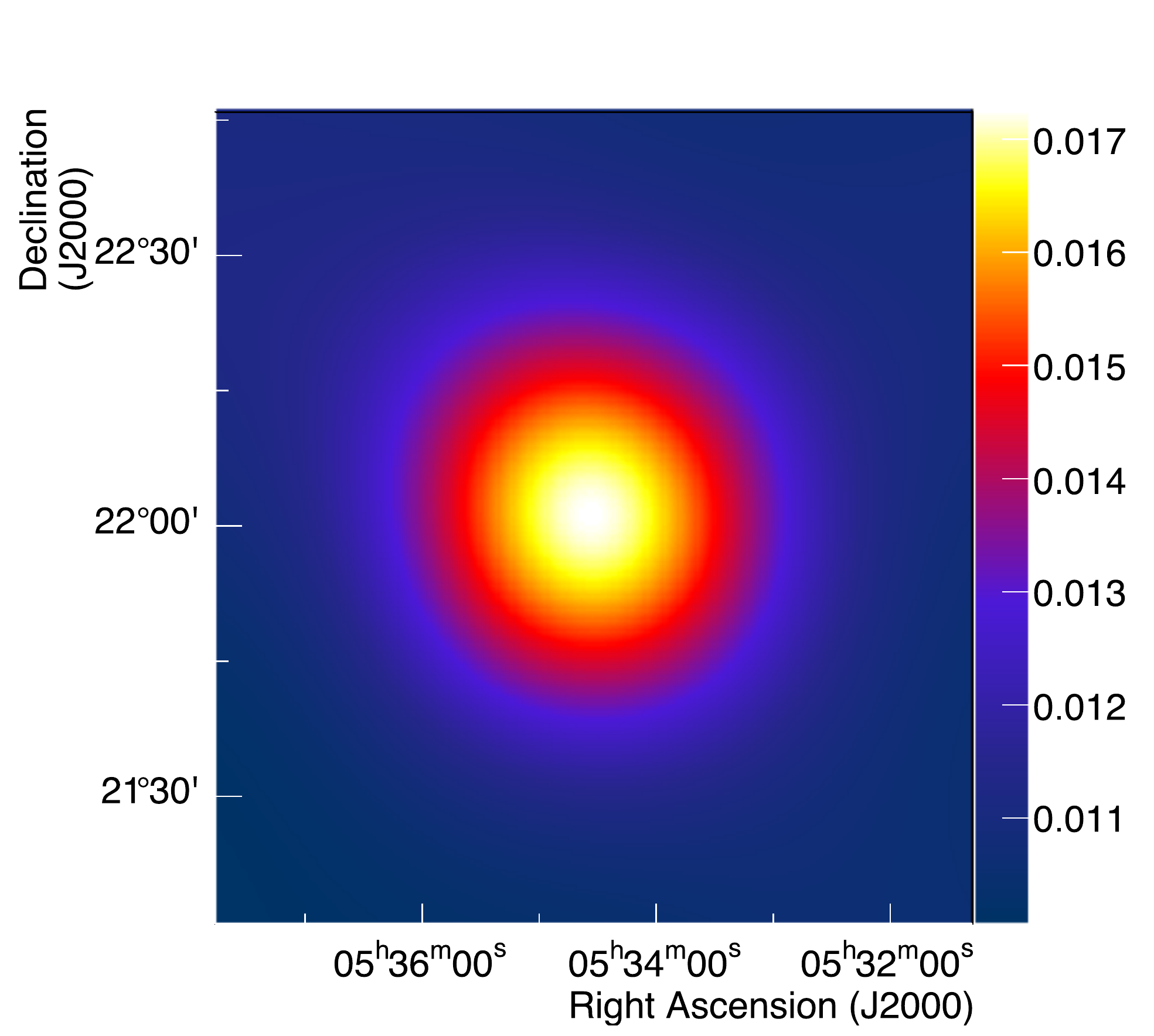}\hfill
	\includegraphics[width=.45\textwidth]{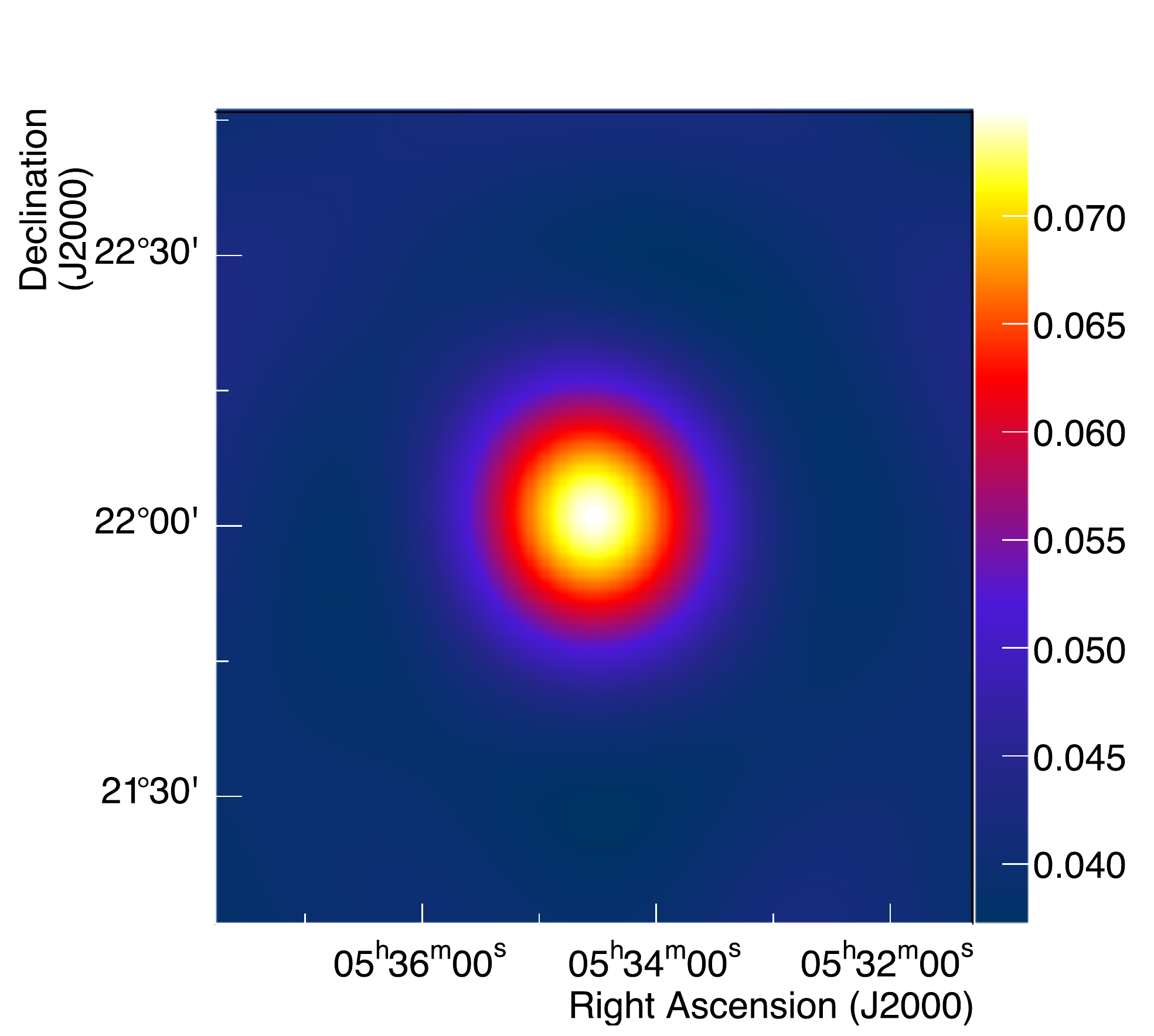}\\
	\includegraphics[width=.45\textwidth]{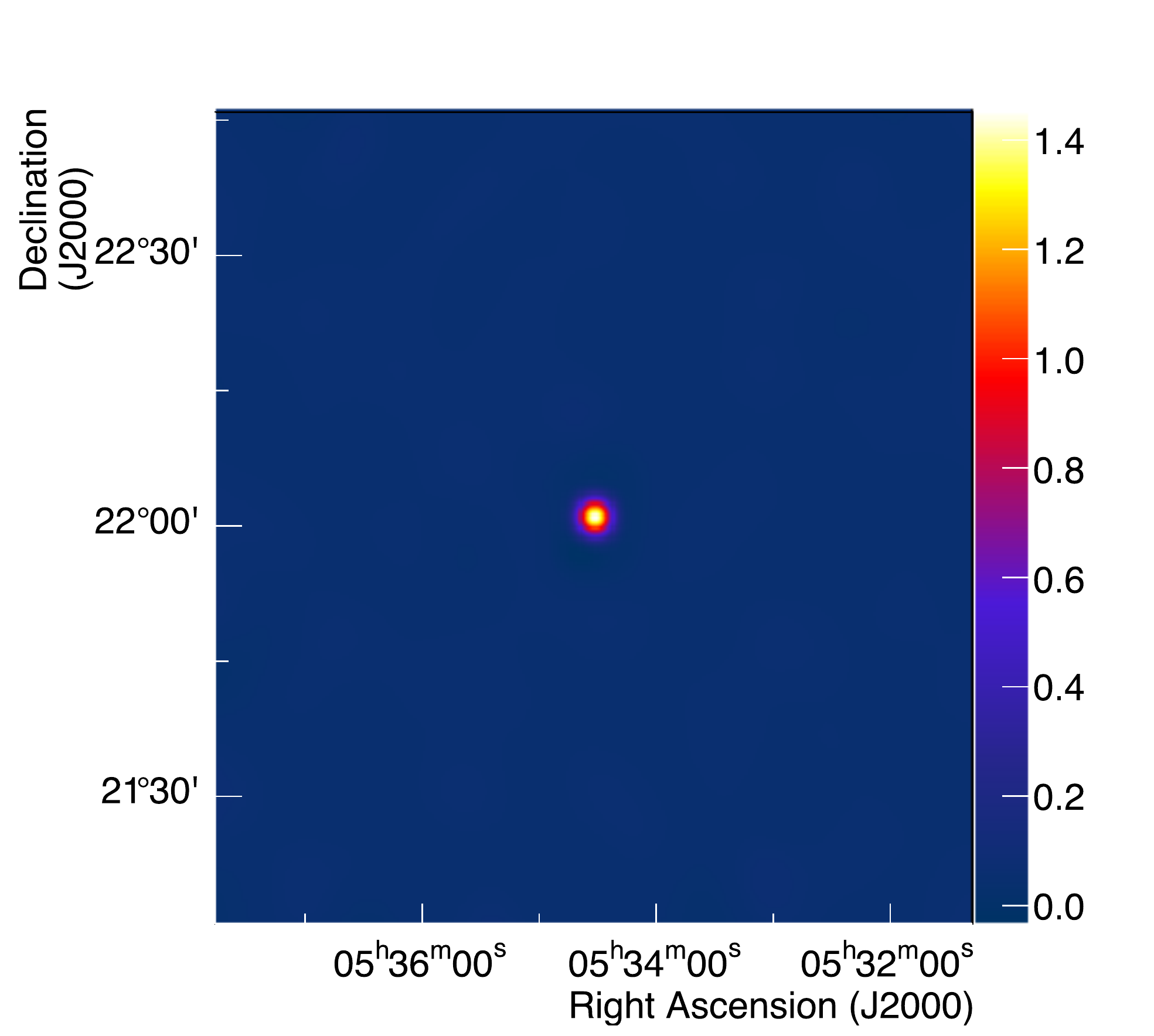}\hfill
	\includegraphics[width=.45\textwidth]{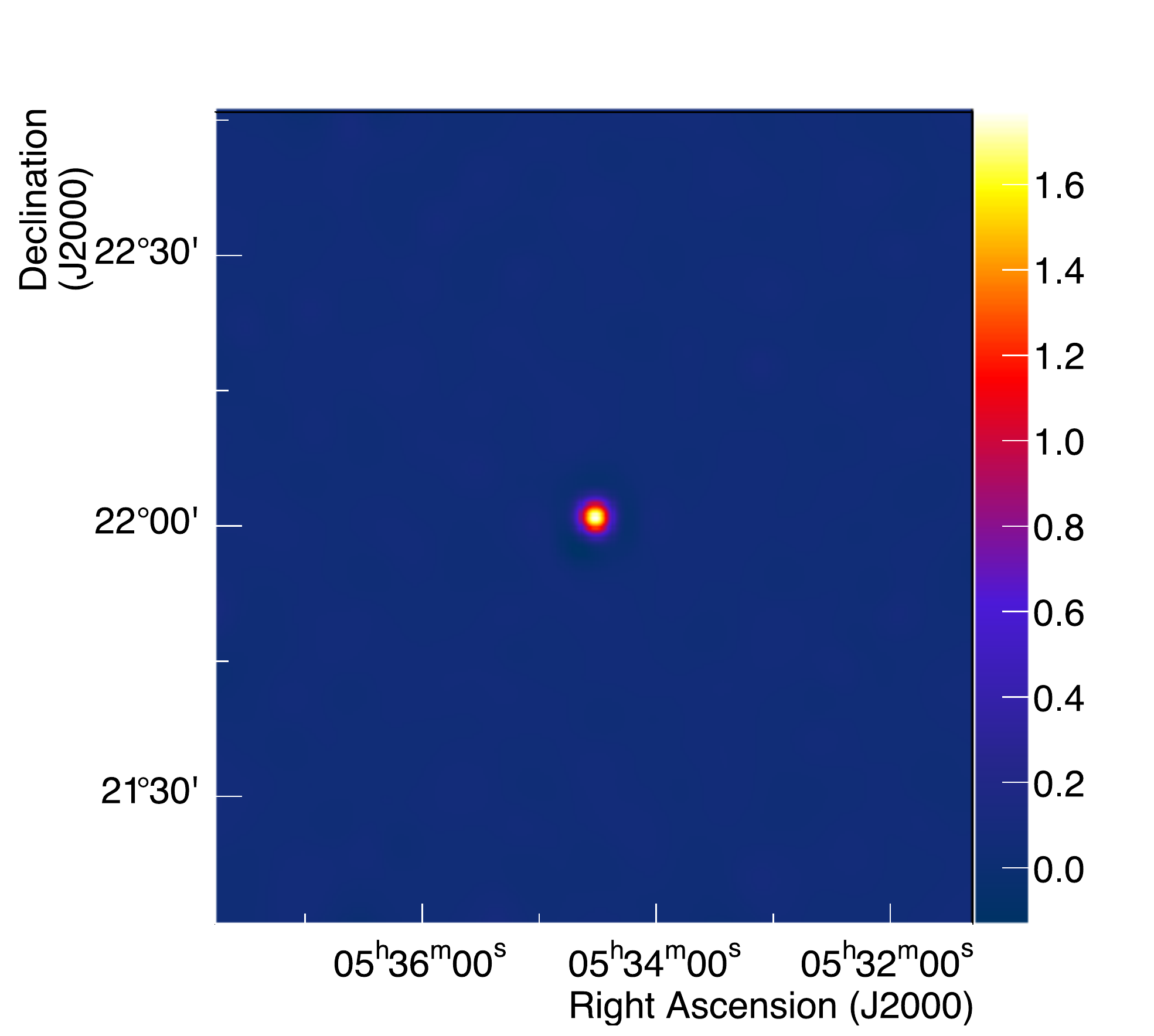}
	\caption{Examples of deconvolved skymaps of a point source with a significance of 50 sigma. \emph{From left to right}: The values for the parameter \texttt{maxiter} are 1, 8, 64 and 1024, respectively.}
	\label{fig:deconv}
\end{figure}

To obtain deep inside into the algorithm and to achieve the best possible results it is essential to understand their influence on the deconvolution. We have performed a detailed study of the influence of the input parameters. Therefore we performed a multidimensional scan with 3168 parameter sets (for the applied steps see  Table~\ref{tab:parameter}). To obtain statistically reliable results
this scan was repeated ten times for different simulations. Examples of deconvolved skymaps can be seen in Figure~\ref{fig:deconv}. As the deconvolution result proved to be robust against varying values of \texttt{icfwidth} only the results of the remaining four parameters are shown in Section~\ref{sec:results}.

\section{Analysis \& Results}
\label{sec:results}

To judge which parameter set yields the best deconvolution results, we follow the example of~\cite{Heinz2012} and take the relative error (cf.\@ Equation~\ref{eq:re}) as a measurement for the goodness of the deconvolution.
\begin{equation}
RE = \sqrt{\frac{\sum\limits_{i,j} \left[ N_{\mathrm{deconvolution}} \left(i, j \right) - N_{\mathrm{model}} \left(i, j \right)\right]^2 }{\sum\limits_{i,j}  \: N_{\mathrm{model}} \left(i, j \right)^2}}
\label{eq:re}
\end{equation}

For the deconvolutions of a point source there is a second method to appraise the results: a Gaussian is fitted to the deconvolved skymaps, the width of the Gaussian is a measure for the angular resolution and can therefore be used as a criterion to find the optimal parameter set.

\begin{figure}[ht]
	\centering
	\includegraphics[trim=1cm 0cm 2.5cm 0.6cm, clip=true,width=.45\textwidth]{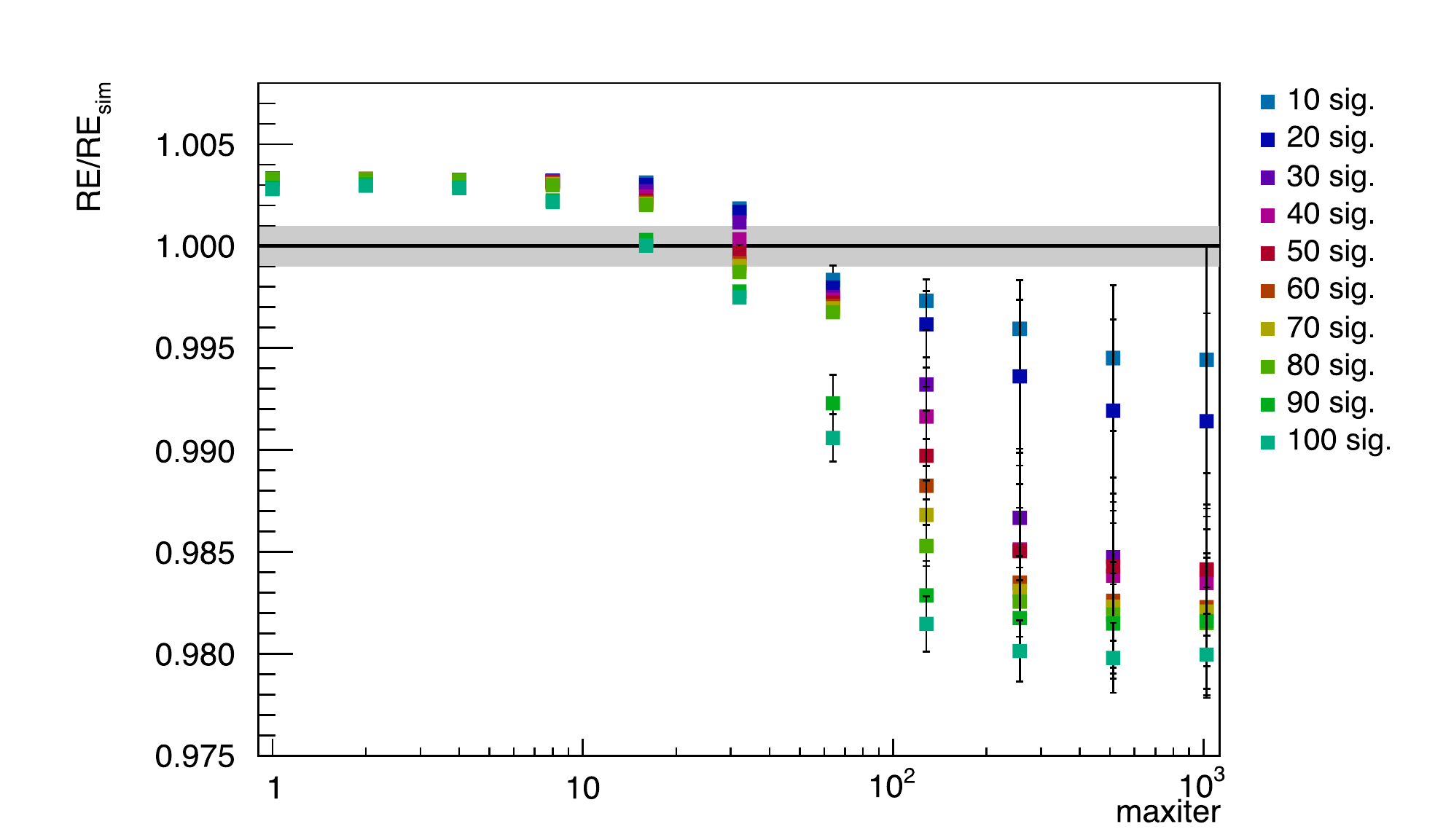}\hfill
	\includegraphics[trim=1cm 0cm 2.5cm 0.6cm, clip=true,width=.45\textwidth]{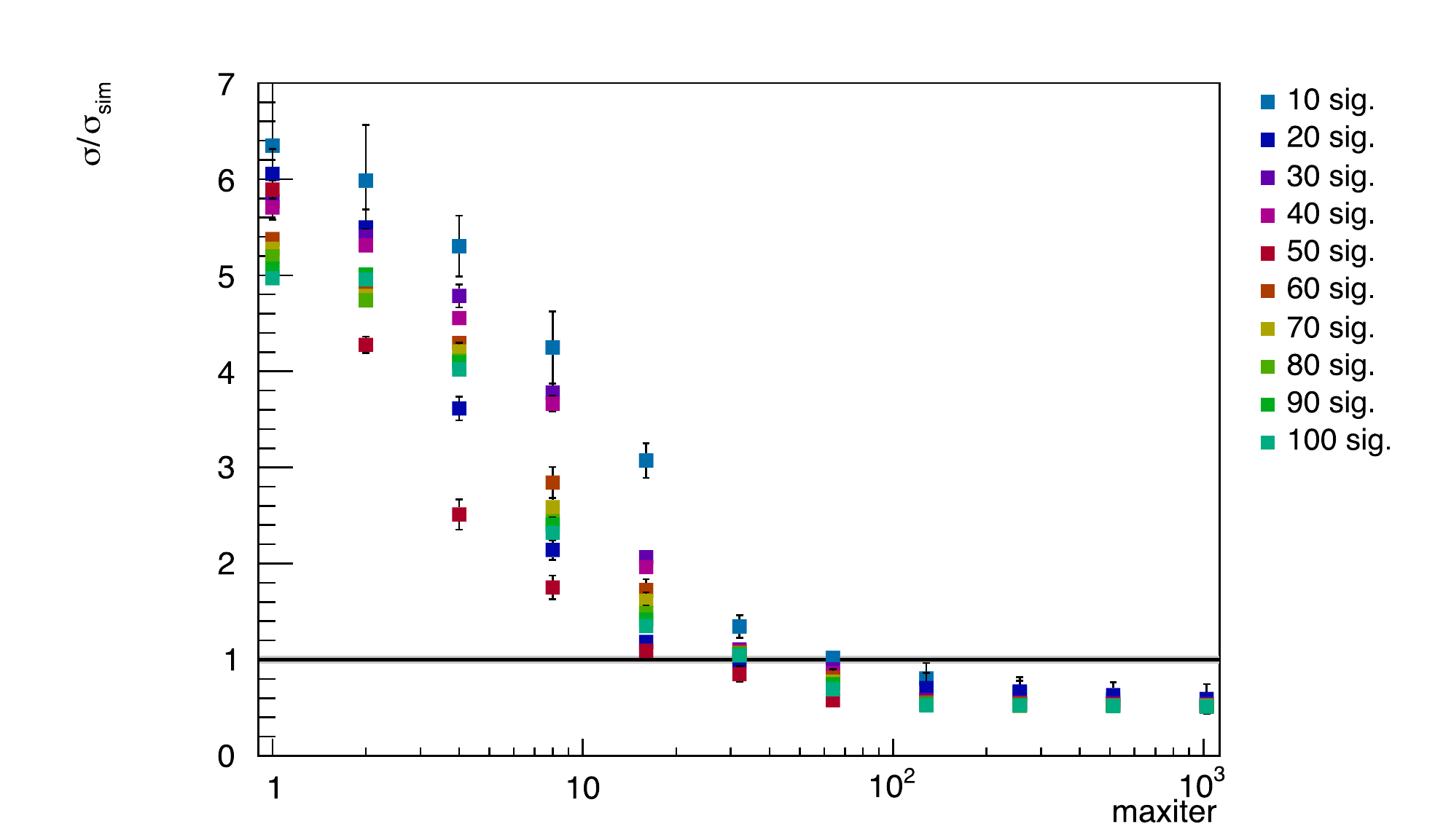}\\
	\includegraphics[trim=1cm 0cm 2.5cm 0.6cm, clip=true,width=.45\textwidth]{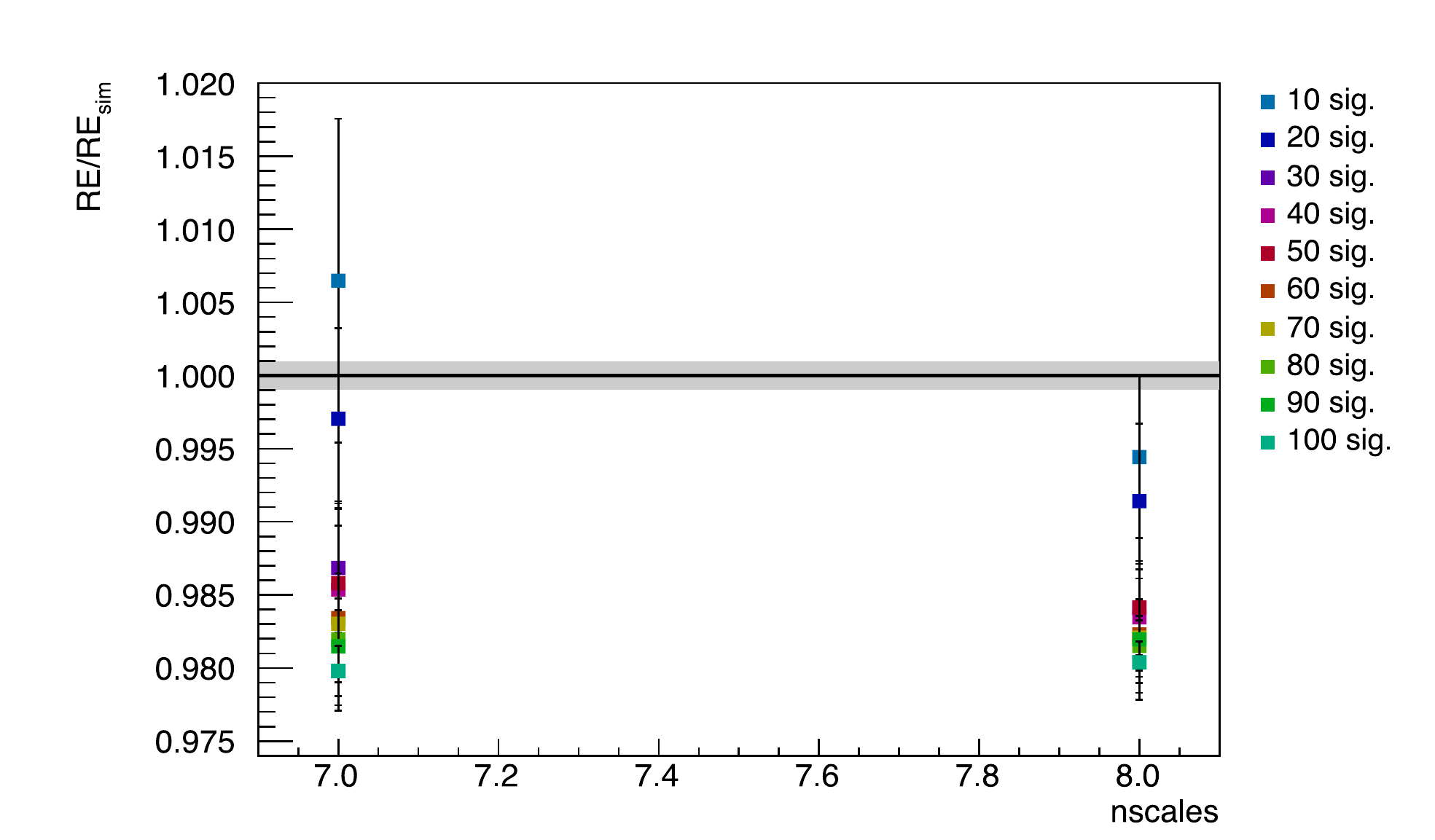}\hfill
	\includegraphics[trim=1cm 0cm 2.5cm 0.6cm, clip=true,width=.45\textwidth]{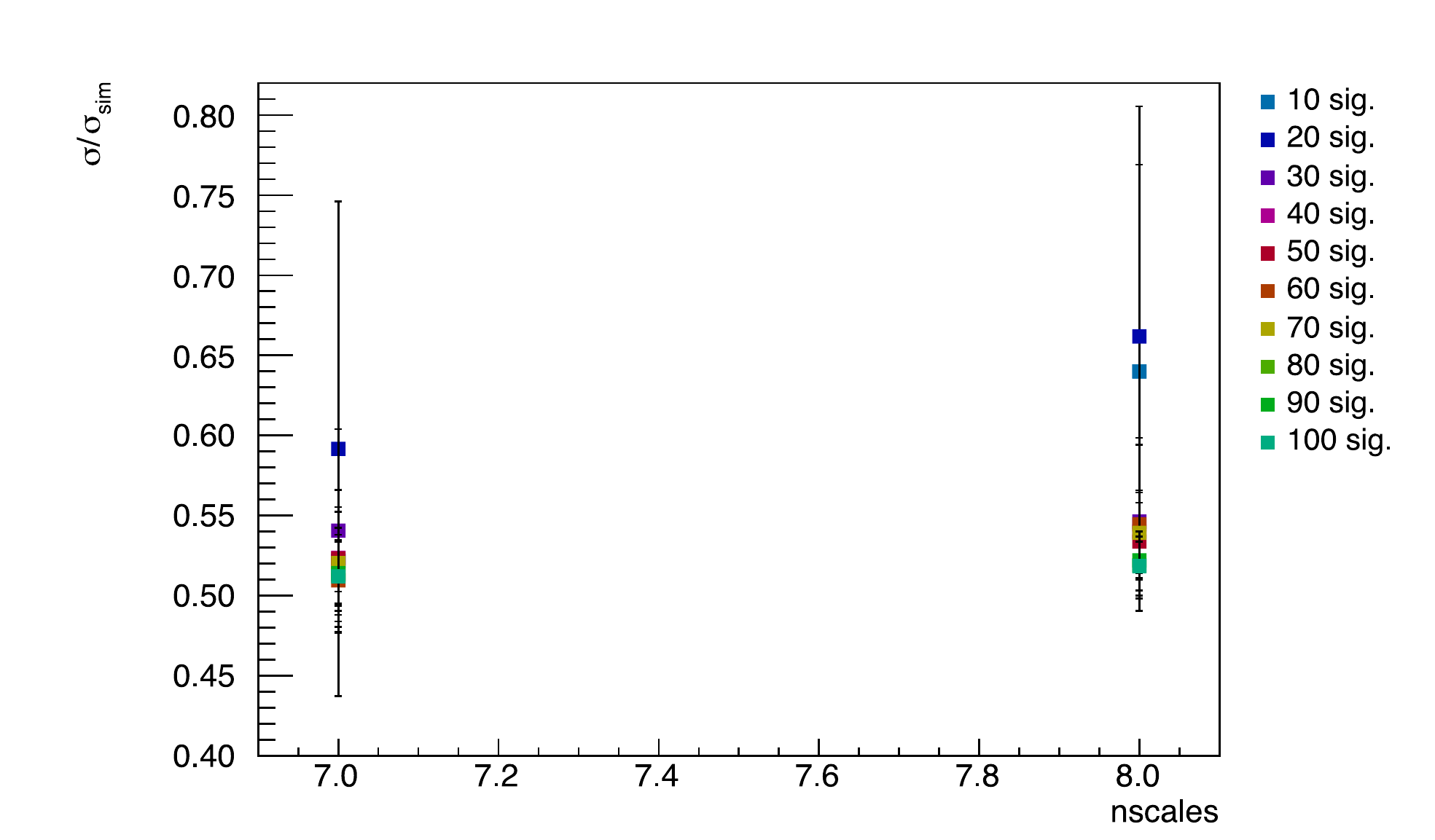}\\	
	\includegraphics[trim=1cm 0cm 2.5cm 0.6cm, clip=true,width=.45\textwidth]{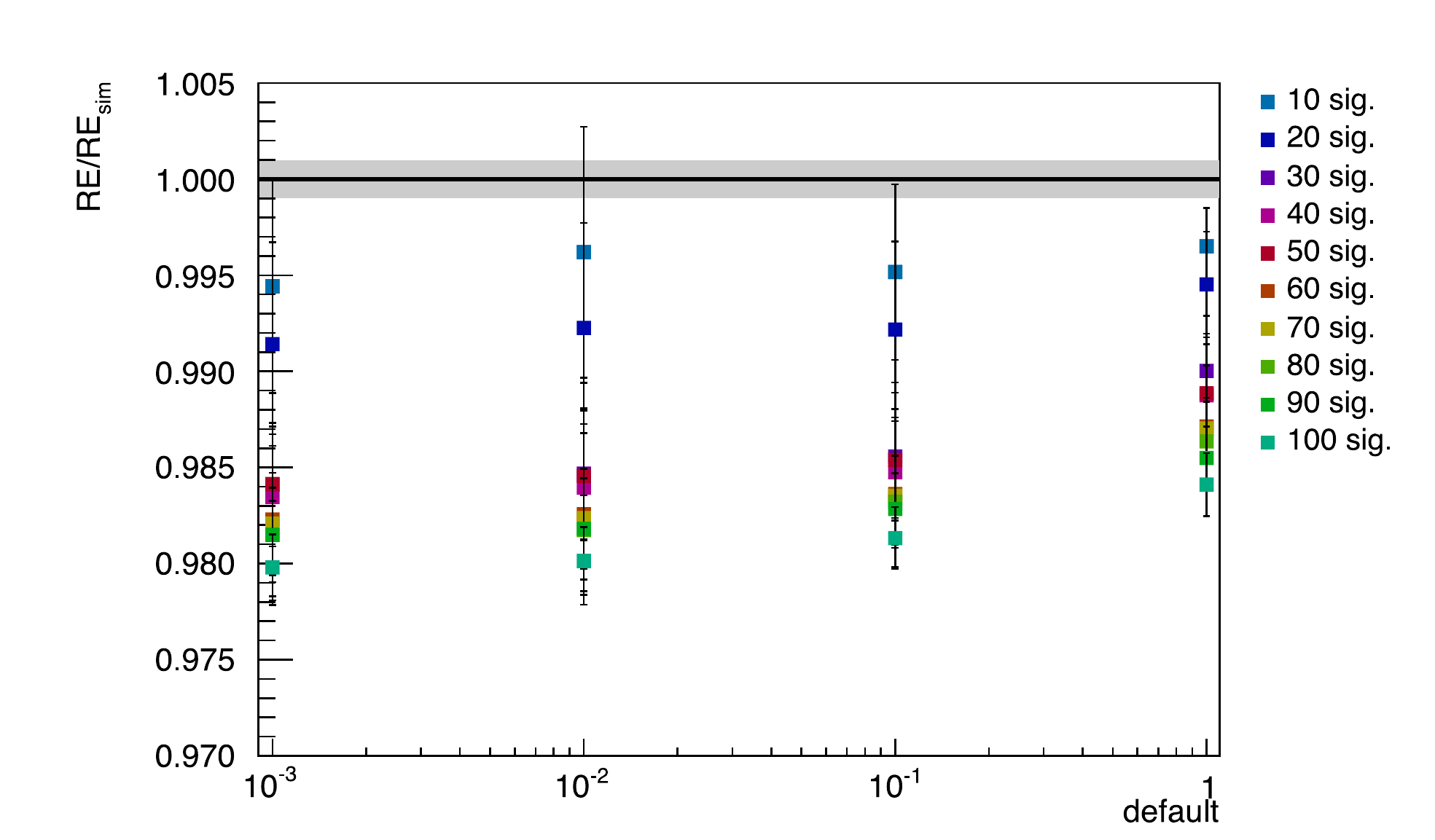}\hfill
	\includegraphics[trim=1cm 0cm 2.5cm 0.6cm, clip=true,width=.45\textwidth]{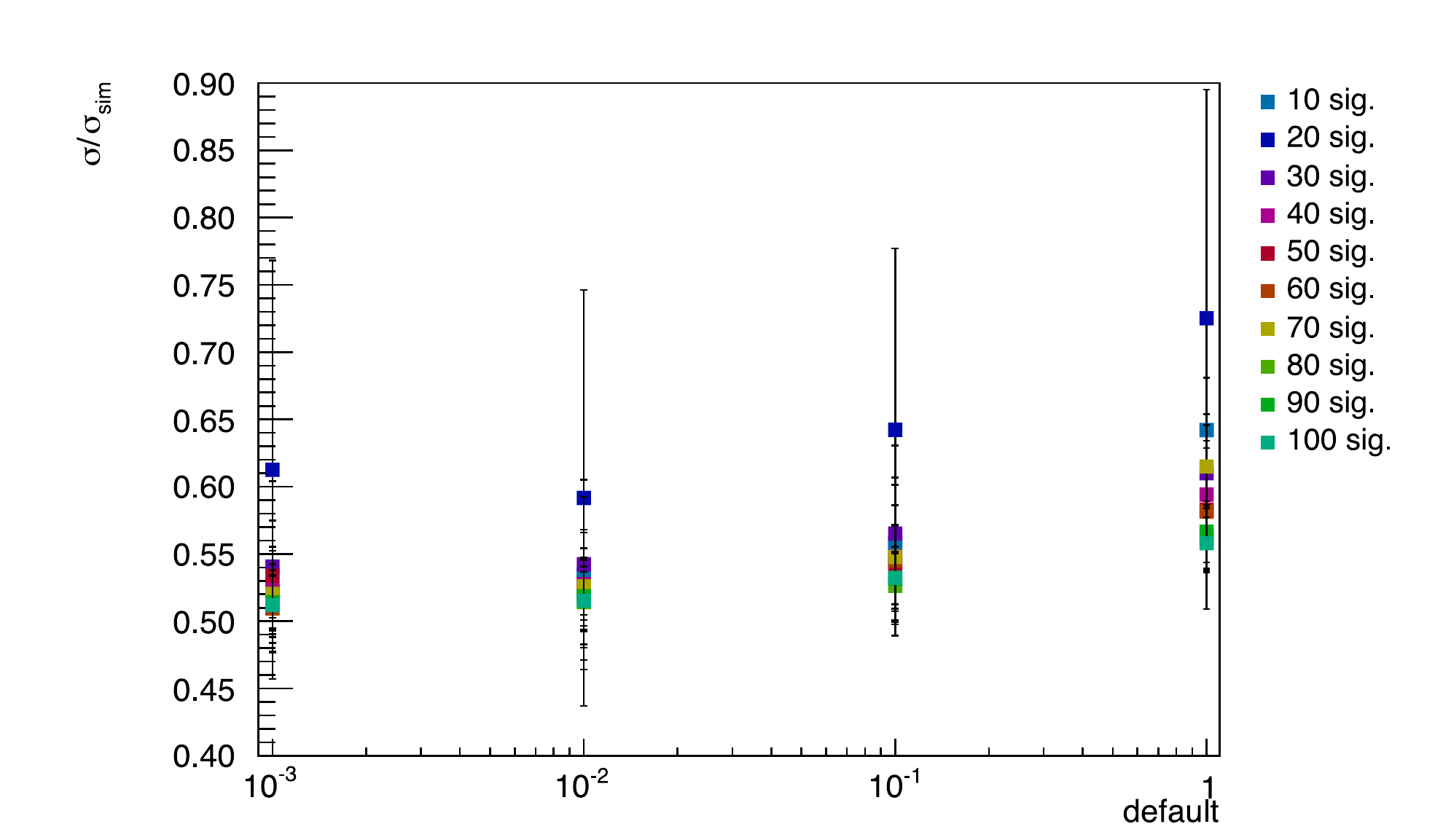}\\
	\includegraphics[trim=1cm 0cm 2.5cm 0.6cm, clip=true,width=.45\textwidth]{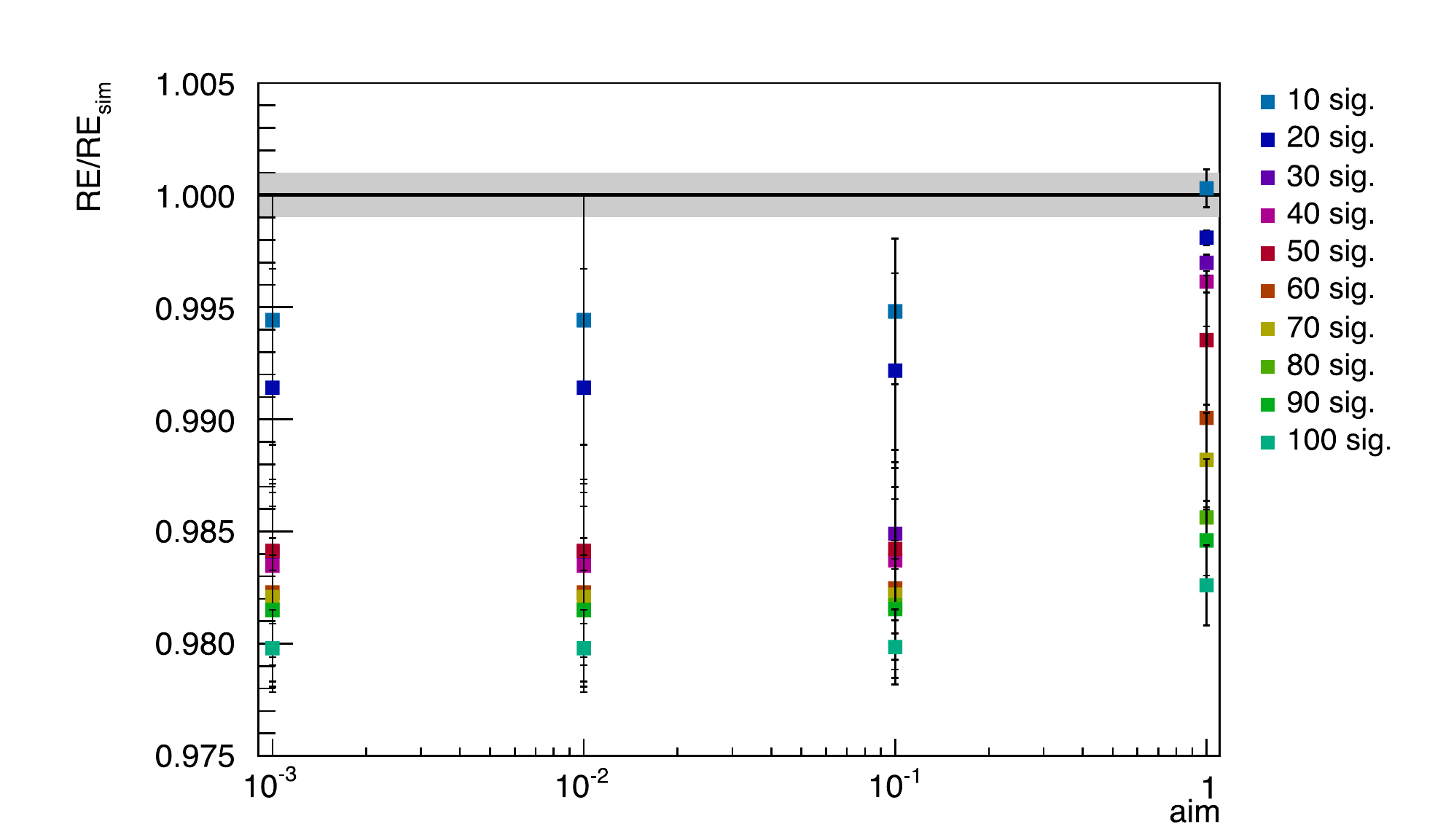}\hfill
	\includegraphics[trim=1cm 0cm 2.5cm 0.6cm, clip=true,width=.45\textwidth]{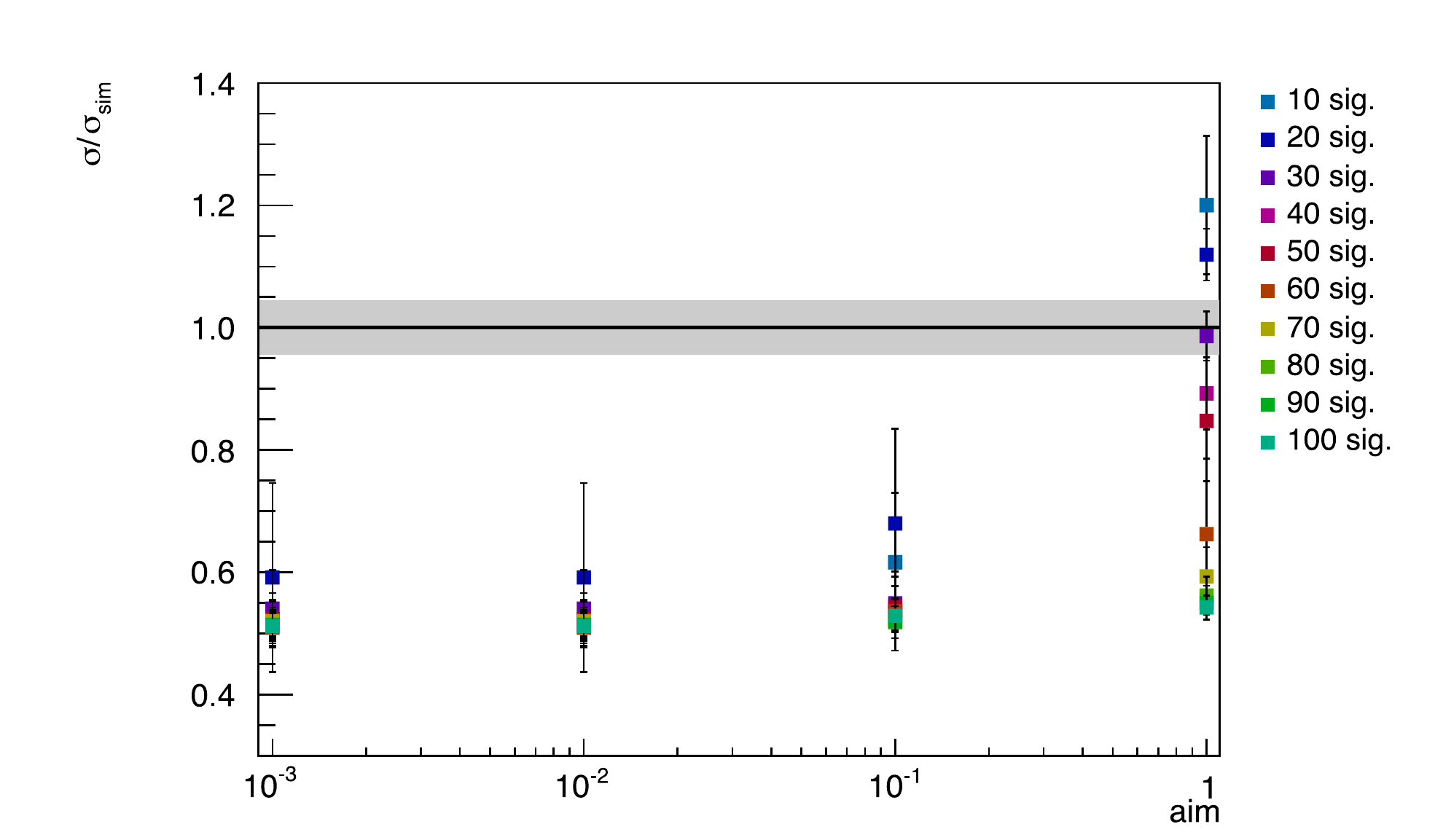}\\
	\includegraphics[width=\textwidth]{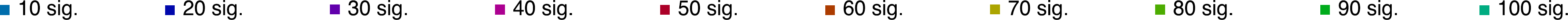}	
	\caption{Results of a systematic scan of deconvolutions for point sources of significances, colour coded from 10 sigma (blue) to 100 sigma (green). From top to bottom, the parameters \texttt{maxiter}, \texttt{nscales}, \texttt{default} and \texttt{aim} are given. In the left hand column the relative error RE normalised to the RE$_{\mathrm{sim}}$ of the undeconvolved simulations is given, on the right hand side the width $\sigma$ of a Gaussian fitted to the deconvolved skymaps, normalised to the width $\sigma_{\mathrm{sim}}$ of the undeconvolved simulations. The grey areas depict the standard deviations of RE$_{\mathrm{sim}}$ and $\sigma_{\mathrm{sim}}$.}
	\label{fig:results}
\end{figure} 

\clearpage
We evaluated the results of parameter scans for point sources with significances between 10 and 100 sigma. The best parameter set, e.g.\@ that yielding the smallest values for the relative error RE and the Gaussian widths, was taken to illustrate the influence of the parameters in Figure~\ref{fig:results}. All parameters were kept fixed at their best values except the one shown in the respective image. On the left hand side the relative errors RE are given, normalised to the relative error RE$_{\mathrm{sim}}$ of the
undeconvolved simulations. On the right hand side the same for the width of a Gaussian fitted to the deconvolved skymaps. 
The errorbars each give the standard deviation from the ten deconvolutions using the same parameter set.

As seen on Figure~\ref{fig:results} (\emph{top panel}) the parameter \texttt{maxiter} has the largest influence and the width of the Gaussian can be improved for values of \texttt{maxiter} > 64 by up to a factor of 1{.}95 compared to the one of the undeconvolved simulations. Larger values might be able to slowly improve the results further, but are accompanied by an approximately linear increase of computing time. The other parameters, illustrated in Figure~\ref{fig:results} (\emph{secondary to bottom panel}), are harder to grasp. Due to the larger error bars no definite statement can be made, but one might perceive some trends. The parameter \texttt{aim} seems to favour smaller values, just like the parameter \texttt{default}. 

What can clearly be seen on all graphs, is that deconvolution works generally better for sources with a higher significance and can improve the angular resolution by a factor up to 1{.}95. Furthermore the optimal parameters tend to be the same for all tested significances.

\section{Conclusion \& Outlook}
\label{sec:sum}

The detailed study, presented here, has shown, that the Maximum Entropy Deconvolution is a feasible choice to improve the angular resolution of VHE $\gamma$-ray skymaps which yield a rather bad signal-to-noise ratio compared to other wavelengths.

The potential improvement in the angular resolution is compatible with the results of a similar study with an independent algorithm \cite{Heinz2012}. In addition it can be asserted, that it is important to understand the influences of various parameters in order to derive a good set of parameters before applying the deconvolution algorithm to real data.

The next logical step will be the systematic study of parameter influences for extended sources of different types and the application to real data.

\end{document}